\documentclass[11pt,preprint]{aastex}

\def\uK{\mu {\rm K}}

\def\deg{^{\circ}}

\def\farcm{\hbox{$.\mkern-4mu^\prime$}}

\def\bu{{\bf u}}

\def\bw{{\bf w}}

\def\by{{\bf y}}

\def\bC{{\bf C}}
\def\bF{{\bf F}}
\def\bI{{\bf I}}
\def\bL{{\bf L}}

\def\bN{{\bf N}}

\def\bS{{\bf S}}
\def\bV{{\bf V}}

\def\tA{{\widetilde{A}}}

\begin{document}

\title{$E/B$ Separation in CMB Interferometry}

\author{Chan-Gyung Park\altaffilmark{1} and Kin-Wang Ng\altaffilmark{2,3}}
\altaffiltext{1}{Korea Institute for Advanced Study, 130-722, Seoul, Korea;
                 parkc@kias.re.kr.}
\altaffiltext{2}{Institute of Physics, Academia Sinica, Taipei, Taiwan 
                 11529, R.O.C.; 
                 nkw@phys.sinica.edu.tw.}
\altaffiltext{3}{Institute of Astronomy and Astrophysics, Academia Sinica, 
                 Taipei, Taiwan 11529, R.O.C.}

\begin{abstract}
We study the problem of separating $E$ and $B$ modes in interferometric 
observations of the polarization of the cosmic microwave background.
The $E$ and $B$ band powers and their mixings are measured from both 
single-dish and interferometric mock observations using the quadratic 
estimator of the maximum likelihood analysis.  
We find that the interferometer can separate $E$ and $B$ modes in a 
single-pointing measurement and is thus well suited for detecting the
faint lensing induced and gravity-wave induced $B$ modes. 
In mosaicking observation, compared to the single dish, the interferometer is 
in general more efficient in separating $E$ and $B$ modes, and for high
signal-to-noise per pixel it needs about three times fewer pixels to measure 
extremely blue polarization power spectra.
\end{abstract}


\section{Introduction} 

Measuring polarization of the cosmic microwave background (CMB) has become
one of the main goals of CMB experiments \citep{sta99,tim02}. 
A CMB polarization field can be decomposed into an electric-type $E$ mode 
and a magnetic-type $B$ mode \citep{zalsel97,kam97}. Recently, the DASI
instrument, a ground-based interferometric array
with degree-scale resolution, has detected the CMB $E$-polarization and $TE$ 
cross-correlation, while setting an upper limit on the $B$-polarization 
\citep{kov02}. 
Most recently, the Wilkinson Microwave Anisotropy Probe has measured the $TE$ 
power spectrum \citep{kog03}, which is consistent with theoretical predictions 
based on the measured CMB anisotropy and indicates a significant 
large-scale $E$-polarization.

The CMB polarization contains a wealth of information about the early universe.
On small scales the $E$-polarization can serve as an independent test of the
physics taking place on the last scattering surface \citep{bon84}, while
on large scales providing a sensitive test of the re-ionization history of 
the universe \citep{ng96,zal97}. 
It can cross-check the measured anisotropy power spectrum and improve 
the accuracy in determining the cosmological parameters \citep{zalspe97}. 
In particular, the $B$ mode is a unique probe of the presence of large-scale 
gravitational waves, a prediction of inflationary models \citep{zalsel97,kam97}.
A detection of the $B$ mode would be essential for reconstructing the inflaton 
potential \citep{lid97}. On smaller scales, measuring secondary $B$ modes 
generated from gravitational lensing could provide information about the 
clustering of cosmic matter \citep{zalsel98}. 
Therefore, the separation of the observed polarization into $E$ and $B$ modes 
is a prerequisite for extracting useful information from data. 

It is clear that there would be no $E$ and $B$ mixing if we have a full-sky
map with infinite resolution. However, for a finite sky patch
observed in a single-dish experiment,
a substantial leakage between $E$ and $B$ was found on large angular scales
when using a quadratic estimator method for measuring the $E$ and $B$ power 
spectra \citep{teg01}. This leakage is caused by ambiguous modes that receive 
contributions to their power from both $E$ and $B$ \citep{lew02,bun03}. 
Furthermore, the effect of aliasing of small-scale power due to the map 
pixelization is another source of leakage, which could be a serious problem 
because the CMB polarization power spectrum is expected to be extremely blue. 
As a result, to achieve the same level of contamination by aliased 
power, one has to oversample the polarization map 2 to 3 times more than the 
temperature map \citep{bun03}. 
In this paper, we revisit the effects of finite size and 
pixelization to $E$ and $B$ mixing but in interferometry experiments. 

\section{CMB Interferometry}

An interferometric array is intrinsically a high-resolution 
polarimetric instrument well suited for observing small-scale polarized 
intensity fluctuations, while being flexible in coverage of a wide range of 
angular scales with the resolution and sensitivity determined by the aperture 
of each element of the array and the baselines formed by the array elements.
Being ground-based, it is controllable and it can track the sky for an 
extensive period of time, as practiced successfully by the DASI team in 
measuring the CMB $E$-polarization \citep{kov02}.
Observational strategies of CMB interferometry experiments such as DASI, CBI,
VSA, and AMiBA can be found in \citet{par03} and references therein.

If the dual-polarization feeds of an interferometer measure the right and left 
circular polarizations, the output will be the four correlations:
$\langle RR^* \rangle$, $\langle RL^* \rangle$,
$\langle LR^* \rangle$, and $\langle LL^* \rangle$.
They can be related to Stokes parameters $(T,Q,U,V)$ by
their associated visibility functions:
$\langle RR^* \rangle = V^T + V^V$,
$\langle LL^* \rangle = V^T - V^V$, 
$\langle RL^* \rangle = V^Q + iV^U \equiv V^+$, and
$\langle LR^* \rangle = V^Q - iV^U \equiv V^-$.
Henceforth, we assume $V=0$ because CMB is expected to have 
no circular polarization.
In typical interferometric measurements, the observation wavelength $\lambda$
is much smaller than the diameter of a dish $D$. So, the sky can be treated as
flat, being spanned by a two-dimensional vector ${\bf x}$. 
Hence, the visibility is the two-dimensional Fourier transform of 
the Stokes parameter multiplied by the primary beam,
\begin{equation}
   V^X_{\bf y}({\bf u}) = b_\nu
   \int d{\bf x} A({\bf x}-{\bf y}) X({\bf x}) e^{2\pi i{\bf u}\cdot {\bf x}},
\end{equation}
where ${\bf u}$ is the two-dimensional projection vector 
(in unit of wavelength) 
of the baseline between two dishes in the ${\bf x}$-plane, ${\bf y}$ is a
pointing position on the sky, $b_\nu$ is a conversion factor from temperature 
to intensity dependent on the observation frequency $\nu$, and $X$ denotes $T$,
$Q$, or $U$ field given by
\begin{eqnarray}
T({\bf x}) &=& \int d{\bf u} 
      {\widetilde T}({\bf u}) e^{-2\pi i{\bf u}\cdot {\bf x}}, \\
Q({\bf x})\pm iU({\bf x}) &=& \int d{\bf u} 
\left[{\widetilde E(\bf u)}\pm i{\widetilde B(\bf u)}\right] 
e^{\pm i2\phi_{\bf u}} e^{-2\pi i{\bf u}\cdot {\bf x}}, 
\end{eqnarray}
where $\phi_{\bf u}$ is the phase in the Fourier space given by
the direction angle of ${\bf u}$,
$\langle {\widetilde Y}({\bf u}){\widetilde Y}^*({\bf w}) \rangle =
S_{YY}(u)\delta({\bf u}-{\bf w})$ ($Y=T,E,B$), and
$\langle {\widetilde T}({\bf u}){\widetilde E}^*({\bf w}) \rangle =
S_{TE}(u)\delta({\bf u}-{\bf w})$.
The power spectrum $S(u)$ defined in the ${\bf u}$-plane can be related
to the angular power spectrum $C_{\ell}$ defined on the sphere by 
$\mathcal{C}_\ell \equiv \ell (\ell+1) C_{\ell} / {2\pi}\approx 2\pi u^2 S(u)$ 
with $\ell \approx 2\pi u$.

\section{$E$ and $B$ Band Powers in Single-pointing and Mosaicking}

\subsection{Power spectrum estimation}

To estimate CMB polarization band powers, we use a quadratic estimator
based on the maximum likelihood analysis, defined as \citep{bon98,par03} 
\begin{equation}
   \delta \mathcal{\widetilde{C}}_b 
    = {1 \over 2} \sum_{b'} (\bF^{-1})_{bb'}
      {\rm Tr} \left[ (\bV\bV^T - \bC) \bC^{-1} 
      {{\partial \bS} \over {\partial\widetilde{\mathcal{C}}_{b'}}} 
      \bC^{-1} \right],
    \label{qe}
\end{equation}
where $\bV$ is the visibility data vector composed of measured $V^+$ and $V^-$ 
quantities, $\bC$ is the sum of signal and noise covariance matrices
($\bC = \bS + \bN$, see below for the definition of a covariance matrix), 
and $\bF$ is the Fisher information matrix defined as
\begin{equation}
F_{bb'} = {1 \over 2} {\rm Tr} 
          \left(\bC^{-1}{{\partial \bS} \over 
	  {\partial \mathcal{\widetilde{C}}_b}}
          \bC^{-1}{{\partial \bS} \over 
	  {\partial \mathcal{\widetilde{C}}_{b'}}} \right).
\label{fishbb}
\end{equation}
By using the Newton-Raphson method, after several iterations we can find 
a set of band powers that most likely fits the data and thus maximizes the 
likelihood function. 
To quantify the sensitivity of each band power to the CMB power spectrum,
we use the band power window functions defined as \citep{kuo02}
\begin{equation}
\widetilde{W}_{b\ell} / \ell = \sum_{b'} (\bF^{-1})_{bb'} \mathcal{F}_{b'\ell},
\label{winfuc}
\end{equation}
where $\mathcal{F}_{b'\ell} \equiv {1 \over 2} {\rm Tr} (\bC^{-1} 
{{\partial \bS} / {\partial \mathcal{\widetilde{C}}_{b'}}}\bC^{-1} 
{{\partial \bS} / {\partial \mathcal{C}_\ell}} )$. 
The window function $\widetilde{W}_{b\ell} / \ell$ 
for the $b$-th band should have a property of the form 
$\left<\mathcal{\widetilde{C}}_b\right> = 
\sum_{\ell}(\widetilde{W}_{b\ell}/\ell)\mathcal{C}_{\ell}$,
where $\mathcal{\widetilde{C}}_b$ is the measured band power from the quadratic 
estimator, and $\sum_{\ell} ( \widetilde{W}_{b\ell}/\ell ) = 1$. 
Those band powers intrinsically bear anti-correlations among 
themselves due to the partial sky coverage and the complex noise property 
of the experiment.
By performing a linear transformation with a transformation matrix as given by
the Hermitian square root of the Fisher matrix $\bF^{1/2}$,
we can get a set of decorrelated band powers and the corresponding window
functions \citep{bon98,teg01},
\begin{eqnarray}
   \mathcal{C}_b &=& \sum_{b'} (\bF^{1/2})_{bb'} \mathcal{\widetilde{C}}_{b'}
                              \Big/ \sum_{b'} (\bF^{1/2})_{bb'}, 
   \label{dcpow} \\
   W_{b\ell} / \ell &=& \sum_{b'}
   (\bF^{-1/2})_{bb'} \mathcal{F}_{b'\ell} \Big/ \sum_{b'} (\bF^{1/2})_{bb'}.
   \label{dcwin}
\end{eqnarray}
The size of the error bar for each decorrelated band power is obtained from
\begin{equation}
   \sigma(\mathcal{C}_b) = 1 \Big/ \sum_{b'} (\bF^{1/2})_{bb'}.
\end{equation}

Following \citet{teg01}, we define $2 \times 2$ leakage matrix $\bL_{b}$
for each band, with components given by
\begin{equation}
   L_{b}^{PP'} \equiv \sum_{\ell_{\rm min}}^{\ell_{\rm max}} 
             W_{b\ell'}^{PP'} / \ell' ,
\end{equation}
where $P$ and $P'$ denote $E$ or $B$ modes, and $\ell_{\rm min}$ 
($\ell_{\rm max}$) is the minimum (maximum) $\ell$-sensitivity limit of 
the experiment considered. 
Here $W_{b\ell}^{EB} / \ell$ is a part of $E$ mode band power window 
function ($W_{b\ell}^{E} / \ell$) that is sensitive to the power leaked 
from $B$ to $E$, $W_{b\ell}^{EE} / \ell$ from $E$ to itself, and likewise 
for other combinations.
If there is no leakage $L_{b}^{EE} = L_{b}^{BB} = 1$ and 
$L_{b}^{EB} =  L_{b}^{BE} = 0$, i.e., $\bL_b = \bI$. As a quantitative measure 
of leakage between $E$ and $B$, we use the ratios of unwanted to wanted 
contributions, i.e., $L^{EB}/L^{EE}$ and $L^{BE}/L^{BB}$ \citep{teg01}.
In the next subsections we discuss the separation of $E$ and $B$ modes 
in each strategy.

\subsection{Single-pointing}

In the {\it single-pointing} strategy, a single field on the sky is tracked for
a long period of time. This is appropriate for an interferometer 
that has uniform ${\bf u}$-coverage and sufficiently small primary 
${\bf u}$-beam size so as to cover a wide range of angular scales 
and resolve the structures in the CMB power spectrum.  
One can reduce sample variances by increasing the number of independent 
fields. This strategy has been adopted by the DASI and the CBI for
deep observation of the CMB \citep{hal02,mas03,kov02}. 

Let us consider a simple 2-element interferometer and choose the 
single pointing position ${\bf y}=0$ without loss of generality, 
then $\langle RL^* \rangle$ and $\langle LR^* \rangle$ are given by
\begin{equation}
V^\pm({\bf u}) = b_\nu
\int d{\bf w} {\widetilde A}({\bf u}-{\bf w}) 
\left[{\widetilde E(\bf w)}\pm i{\widetilde B(\bf w)}\right] e^{\pm i2\phi_{\bf w}},
\end{equation}
where $A({\bf x})=\int d {\bf u} {\widetilde A}({\bf u})
e^{-2\pi i{\bf u}\cdot {\bf x}}$ is assumed to be a 
symmetrical flat-illuminated feed horn,
${\widetilde A}({\bf u}) = (8a^2/\pi^2)(\arccos b - b\sqrt{1-b^2})$,
where $a=\lambda/D$ and $b=u\lambda/D$.
The ensemble averages are a set of simultaneous integral equations 
for $E$ and $B$ power spectra (visibility covariance matrices):
\begin{equation}
\langle V^+({\bf u}) V^{\pm*}({\bf u}) \rangle =  b_\nu^2 \int d{\bf w} 
|{\widetilde A}({\bf u}-{\bf w}) |^2 [S_{EE}(w)\pm S_{BB}(w)]
{1 \choose e^{i4\phi_{\bf w}}}.
\label{enavu}
\end{equation}
For a single-dish experiment, it gives ensemble averages 
$\langle V^+V^{\pm*}\rangle$ simply given by equation~(\ref{enavu}) 
with ${\bf u}=0$. Because of the symmetrical ${\widetilde A}({\bf u})$,
$\langle V^+V^{-*}\rangle=0$.
As such, the separation of $E$ and $B$ is undetermined
although $\langle V^+V^{+*}\rangle$  measures the total polarization power. 
One should be cautious 
about the integration in $\langle V^+V^{\pm*}\rangle$ for small ${\bf w}$ 
(or low $\ell$) where the flat-sky approximation is no longer reliable 
\citep{ng01}. However, full-sky two-point polarization correlation 
functions have been constructed \citep{kam97,ng99}, showing that the 
two-point correlation functions at zero lag, which are equivalent to 
$\langle V^+V^{\pm*}\rangle$, have similar properties.
On the other hand, $\langle V^+({\bf u}) V^{-*}({\bf u}) \rangle$ 
is non-vanishing in general for an interferometer, 
since a narrow range of $\phi_{\bf w}$ can be sampled up to an uncertainty 
of the size of the dish. The information contained in it,
which is orthogonal to $\langle V^+({\bf u}) V^{+*}({\bf u}) \rangle$, 
allows one to separate $E$ and $B$ powers in power spectrum estimation. 

Figure 1 shows schematic diagrams showing different nature of Fourier 
mode samplings in the $\bu$-space by the primary beam patterns 
for the single dish and the interferometer. Big circles represent 
the sensitivity ranges set by the primary beams in the single-pointing
strategy while the small circles denote the narrow synthesized beams that
sample the Fourier modes at particular locations in the mosaicking 
strategy. In the single-pointing strategy, a single dish intrinsically 
has a difficulty in sampling the phase due to the beam centered at the origin,
which makes $E/B$ separation impossible. 
On the other hand, the interferometer is more efficient in phase sampling
because its beam is off-centered (big circles in Fig. 1).
Thus the interferometer has better performance in separating $E$ and $B$ modes.

We can also understand this by considering an ideal case 
in which $u\lambda \gg D$, 
then ${\widetilde A}({\bf u}-{\bf w}) \simeq \delta({\bf u}-{\bf w})$
and hence equation~(\ref{enavu}) becomes
$\langle V^+({\bf u}) V^{\pm*}({\bf u}) \rangle \propto 
[S_{EE}(u)\pm S_{BB}(u)] {1 \choose e^{i4\phi_{\bf u}}}$
from which $E$ and $B$ modes can be completely separated.
To study the dish finite-size effect, we have performed a simulation 
of 60 independent single-pointing CMB observations using 
a 2-element interferometer operating at 95 GHz, with dishes of $D=20$ cm 
separated by 60 cm. 
In generating CMB fields, we simply assume a sine-shape power spectrum 
with a bump at $\ell \approx 1300$ and with $B$ mode having the same 
amplitude of $E$ mode, and adopt an instrumental noise level of 3 $\mu$K per
visibility. 
We have measured the decorrelated band powers from the maximum likelihood 
analysis using equations (\ref{qe}) and (\ref{dcpow}).
The decorrelated band powers and the corresponding window 
functions defined in equation (\ref{dcwin}) are shown in Figure 2.
The $W_{b\ell}^E / \ell$ (2nd row panels) expresses contribution 
to the $b$-th $E$ mode band power from $E$ mode itself (left; 
$W_{b\ell}^{EE} / \ell$) and from $B$ mode powers (right panel; 
$W_{b\ell}^{EB} / \ell$), and likewise for $W_{b\ell}^B / \ell$
(3rd row panels). 
Since the $\bu$-beam with $\Delta u_{\rm FWHM} = 50.7$
is sufficiently narrow and at a rather long baseline of $u=253.3$ 
from the origin, $\phi_\bw$ can be sampled with a small error 
$\delta \phi_\bw /2\pi \simeq 0.03$. The band power window function shows 
that the $E$ and $B$ modes have been cleanly separated with a leakage ratio
of only $L^{EB}/L^{EE} = L^{BE}/L^{BB} \simeq 0.02$ as is expected.

We have also investigated the extraction of the lensing induced $B$ mode from 
CMB polarization observations made by for example the forthcoming 
AMiBA experiment\footnote{http://amiba.asiaa.sinica.edu.tw/} \citep{lo00}, 
which is an interferometric array of 7 elements that are hexagonally 
close-packed dishes of $D=60$ cm operating at 95 GHz with two 10 GHz channels
and a system temperature of $70$ K.
We have used the CMBFAST code \citep{sel96} to generate the scalar-induced
$E$ power spectrum and the lensing induced $B$ power spectrum in 
a flat reionized $\Lambda$CDM cosmological model with model parameters 
$\Omega=1.0$, $\Omega_{\Lambda}=0.73$, $\Omega_b = 0.045$, $h=0.70$, 
$n_s =1.0$, and $\tau_{re}=0.17$. In a single-pointing observation made by the 
7-element AMiBA for an integration time of one year, if $E$ and $B$ modes can be
cleanly separated, the signal-to-noise ratio (excluding sample variance) 
for the $E$ polarization will be about 25, while that 
for the lensing induced $B$ polarization will be about 2.
Figure 3 shows the measured decorrelated band powers and the corresponding 
window functions. We have found rather high leakages from $B$ to $E$ which 
are given by $L^{EB}/L^{EE} \simeq 0.30$ ($0.11$) for the first (second) band. 
However, the lensing induced $B$ power spectrum is quite small such that the 
leakages do not make significant contamination to the measurement of the $E$ 
mode.
On the contrary, although the leakages from $E$ to $B$ are only
$L^{BE}/L^{BB} \simeq 0.05$ ($0.04$) for the first (second) band, 
the effective leakage power from $E$ to $B$ is at about 1 $\mu$K level,
much higher than the genuine lensing induced $B$ power spectrum. 
In order to measure the lensing induced $B$ mode, 
one has to reduce $L^{BE}/L^{BB}$ to less than 0.01. This can be done by
reducing the size of the dish (but then it will decrease the sensitivity) or
using the method of mosaicking as described below.

\subsection{Mosaicking}

In the single-pointing strategy, the resolution in the ${\bf u}$-space 
is limited by the area of the surveyed sky which is equal to the size of 
the primary beam. By {\it mosaicking} several contiguous pointings, 
we can increase the resolution in the ${\bf u}$-space. 
This strategy is essential for close-packed interferometers such as the AMiBA 
and the CBI with ${\bf u}$-beam size larger than the structure of the 
CMB power spectrum \citep{par03}. 

For a pair of pointing positions $\by_m$ and $\by_n$, 
equation~(\ref{enavu}) change into the visibility covariance matrices
\begin{equation}
\langle V_{{\bf y}_m}^+({\bf u}) V_{{\bf y}_n}^{\pm*}({\bf u}) \rangle =  
   b_\nu^2 \int d{\bf w} |{\widetilde A}({\bf u}-{\bf w}) |^2 
   e^{2\pi i ({\bf u}-{\bf w})\cdot ({\bf y}_m - {\bf y}_n)}
   [S_{EE}(w)\pm S_{BB}(w)] {1 \choose e^{i4\phi_{\bf w}}}
\label{enavum}
\end{equation}
and the single-dish covariance matrices
$\langle V_{{\bf y}_m}^+V_{{\bf y}_n}^{\pm*}\rangle$ as  
given by equation~(\ref{enavum}) with ${\bf u}=0$.
Here we see additional phases as a function of $\by_m-\by_n$ 
which can increase the resolution in $\bu$-space by mosaicking neighboring
pointings, as illustrated in Figure 4. Note that the interferometric primary 
beam actually samples the $\bu$-space peaked at $u_x\simeq 206$ which is the 
length of the baseline, and that we have chosen particular numbers and 
separations of pointings to produce similar effective beams. As shown 
in Figure 1, the phases sampled are more reliable in mosaicking 
strategy due to the narrow synthesized beam in the $\bu$-space (small circles).

Figure 5 shows an example of the $E$ and $B$ polarization band power 
estimates expected in a 61-pointing hexagonal mosaic mock observation 
with pointing separation $\delta\theta_{\rm mo} = 6\farcm25$
by the single dish with $D=125$ cm ($\nu = 95$ GHz). 
We have used the same sine-shape power spectrum in generating CMB polarization 
fields and made 60 independent mosaics to reduce the sample variance. 
In order to study the sample-variance limited regime,
we have adopted extremely low instrument noise, $10^{-3}$ $\uK$ 
per pointing, to suppress the complications set in from noise. 
The total survey size per mosaic is about $1\deg$, thus the band width is chosen
to be $\Delta\ell \simeq 500$. The decorrelated band power window functions 
are also shown in the bottom panels of Figure 5. 
The mixing between $E$ and $B$ is significant at the first and last bands 
denoted respectively by the solid and dot-dashed curves.
Furthermore, both band power window functions at the first band,
$W_{b\ell}^{EE}$ and $W_{b\ell}^{EB}$, are rising up steeply in the direction
of $\ell \approx 0$, which reflects that the single-dish instrument
cannot determine the phases $\phi_{\bw}$ near the origin in the $\bu$-space.

Similarly, the polarization band powers and the corresponding band power 
window functions expected in a 19-pointing hexagonal mosaic mock observation
by the 2-element interferometer with $D=60$ cm and 5 cm separation 
($\nu = 95$ GHz) are shown in Figure 6. 
In this case $\delta\theta_{\rm mo} = 10\arcmin$ is chosen in order to make 
the same survey size of $1\deg$, and the instrument noise is again 
assumed to be $10^{-3}$ $\uK$ per visibility.
Unlike the case of single dish in Figure 5, there is no steep rise-up of 
the window functions near $\ell \approx 0$.
The $E$ and $B$ mixing at the first band is still large but smaller than 
that of the single-dish experiment, since it is just as difficult 
to determine $\phi_{\bw}$ at the first band. However, this band is already 
at the tail of the $\ell$-range sensitivity of the interferometer. 
On the contrary, a clean $E/B$ separation can be done at high $\ell$ 
region where a better sampling of $\phi_{\bw}$ is possible.
Note that the interferometer has a much better performance at the last band 
than the single dish mainly due to a higher sensitivity at that band.

Examples of mosaicking given above are too ideal to look into the practical
issues. Cases for realistic instrument noises and CMB power spectra should 
be investigated. 
Figures 7 and 8 show results in the case of realistic instrument noise. 
We have assumed a noise level of 3 $\mu$K per pointing or 
visibility while the sine-shape power spectrum is still used.
Compared to the single dish, the interferometer gives $E$ and $B$ band power
measurements with very large uncertainties. 
The interferometer considered here has about three times fewer pixels
(19-pt) than the single dish does (61-pt) while the same noise level per pixel 
is assumed in both cases. Therefore, the overall noise level is quite higher
in the interferometry.
In spite of this higher noise level, the interferometer
has better performance in separating $E$ and $B$ modes. 
For $E$-estimates in the single-dish experiment (interferometry), 
the ratios of unwanted to wanted modes at the first and the last bands 
are $L^{EB}/L^{EE}=0.50$ ($0.26$) and $0.28$ ($0.04$), respectively.

Figures 9 and 10 also show similar results in the case of realistic instrument
noise but with $\Lambda$CDM power spectrum with zero $B$ mode power.
The noise level is assumed to be 3 $\mu$K per pixel. 
We use the CMBFAST power spectrum \citep{sel96}
of a flat $\Lambda$CDM cosmological model with model parameters 
$\Omega=1.0$, $\Omega_{\Lambda}=0.73$, $\Omega_b = 0.045$, $h=0.70$, 
and $n_s =1.0$. Reionization has not been considered. 
The distinctive feature in the band power estimates is that at the first 
band a significant power leakage from $E$ to $B$ modes is seen, especially 
in the single-dish experiment (Fig. 9; $L^{EB}/L^{EE}=0.57$ and 
$L^{BE}/L^{BB}=0.44$). 
On the other hand, such leakage is relatively smaller in the interferometry
($L^{EB}/L^{EE}=0.31$ and $L^{BE}/L^{BB}=0.25$).
In other words, in a situation of asymmetric shapes of power spectra
between $E$ and $B$ modes (e.g., $\Lambda$CDM case), one mode can be 
significantly contaminated by the other mode, especially in the 
single-dish experiment. 

Figure 11 summarizes the level of $E/B$ mixing in terms of the ratios 
$L^{EB}/L^{EE}$ and $L^{BE}/L^{BB}$ for the single dish and the interferometer.
Here Figures 11$a$ and 11$b$ correspond respectively to the cases of the 
sine-shape power spectrum with negligible noise (Figs. 5 and 6) and 
realistic noise (Figs. 7 and 8), while Figure 11$c$ corresponds to that of the 
$\Lambda$CDM power spectrum and realistic noise (Figs. 9 and 10).
The leakage drops with $\ell$ for the interferometer. 
However, for the single dish the leakage increases at higher
$\ell$ due to $E/B$ mixing caused by sparse sampling of data points. 
This $E/B$ mixing can be removed by the dense data sampling during mosaicking
or map-making process. On the other hand, the interferometer, even with 
sparse data sampling scheme, does not experience large $E/B$ mixing. 
In Figures 11$a$ and 11$b$, the leakage pattern from $B$ to $E$ is essentially 
equal to that from $E$ to $B$ because the $E$ and $B$ power spectra assumed 
are symmetric. However, there exists an effect of asymmetry between the $E$ and 
$B$ mode power spectra in the $\Lambda$CDM case (Fig. 11$c$), where 
$L^{BE}/L^{BB}$ has a minus sign at the second band for the single-dish 
experiment. It should be also noted that at the last band both experiments 
have larger leakage ratios for the case of negligible noise (Fig. 11$a$) 
than those for the case of realistic noise (Fig. 11$b$). This is because 
the window functions for negligible noise have very small sidelobes, 
while those for realistic noise have larger sidelobes fluctuating at the zero 
point and self-canceling, especially at the last band.  

\section{Conclusion}

We have pointed out that the separation of $E$ and $B$ modes in 
CMB measurements depends on how well the phase information of the CMB 
polarization in the Fourier space is obtained. 
This has been shown by performing maximum-likelihood estimations of the band 
powers from mock single-dish data in comparison with that 
from an interferometer. Being able to sample a narrow range of the phase 
by an individual pointing, the interferometer can separate $E$ and $B$ modes
in a single-pointing measurement. This makes interferometry particularly
attractive in the detection of extremely weak $B$ polarization signals
such as the lensing induced $B$ mode and the gravity-wave induced $B$ mode.
In mosaicking observation, the interferometer is generally
more efficient than the single dish in separating $E$ and $B$ modes.
In the sample-variance limited cases, we have found that the 
interferometer needs about three times fewer pixels than the single dish 
to achieve the same or even better measurement of the band powers. 
The main reason is that it has the most sensitivity 
at the peak of the power spectrum. 
Apart from a serious leakage between $E$ and $B$ modes at low $\ell$ region 
due to poor determination of the phase, 
$E$ and $B$ can be separated within the resolution 
limit with a mixing of less than 10 percents. This small mixing arises because
the sidelobes of the effective beam in mosaicking lead to aliased power, or
equivalently increase the uncertainty of sampling the phase. As is known,
this effect can be alleviated by making a dense mosaicked map~\citep{bun03}.
With regard to a similar problem in the field of weak gravitational lensing
that one wants to separate the $E$-mode cosmic shear from other $B$-mode
signals due to intrinsic galaxy alignment or systematic errors 
as discussed in \citet{pad02} and references therein,
we may consider a phase modulated annulus filter analogous to an 
interferometric beam to make a pixelized map from lensing data. 
The work along this line is under progress.

\acknowledgments
We acknowledge valuable discussion with Keiichi Umetsu.
CGP was supported by the BK21 program of the Korean Government
and the Basic Research Program of the Korea Science and Engineering Foundation
(grant 1999-2-113-001-5), and acknowledges the support by the CosPA Center
at the National Taiwan University during his visit.
KWN was supported by the Taiwan NSC Grant NSC92-2112-M-001-029.
Some of the results in this paper have been derived using the CMBFAST package.

\clearpage

\begin{figure}
\epsscale{1.5}
\resizebox{\textwidth}{!}{\includegraphics{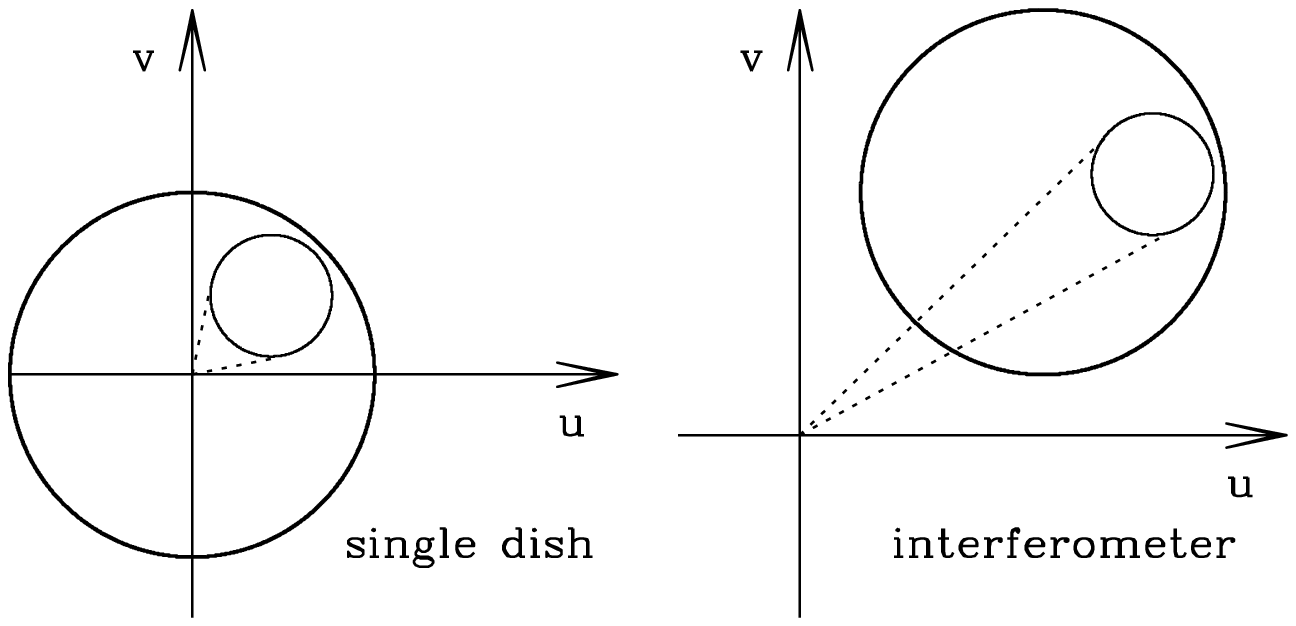}}
\caption{
Schematic diagrams showing different nature of Fourier mode samplings
in the $\bu$-space by the primary beam patterns in the single-dish experiment
(left) and the interferometry (right) for the single-pointing (large circles)
and the mosaicking (small circles).}
\end{figure}
\clearpage

\begin{figure}
\epsscale{1.5}
\resizebox{\textwidth}{!}{\includegraphics{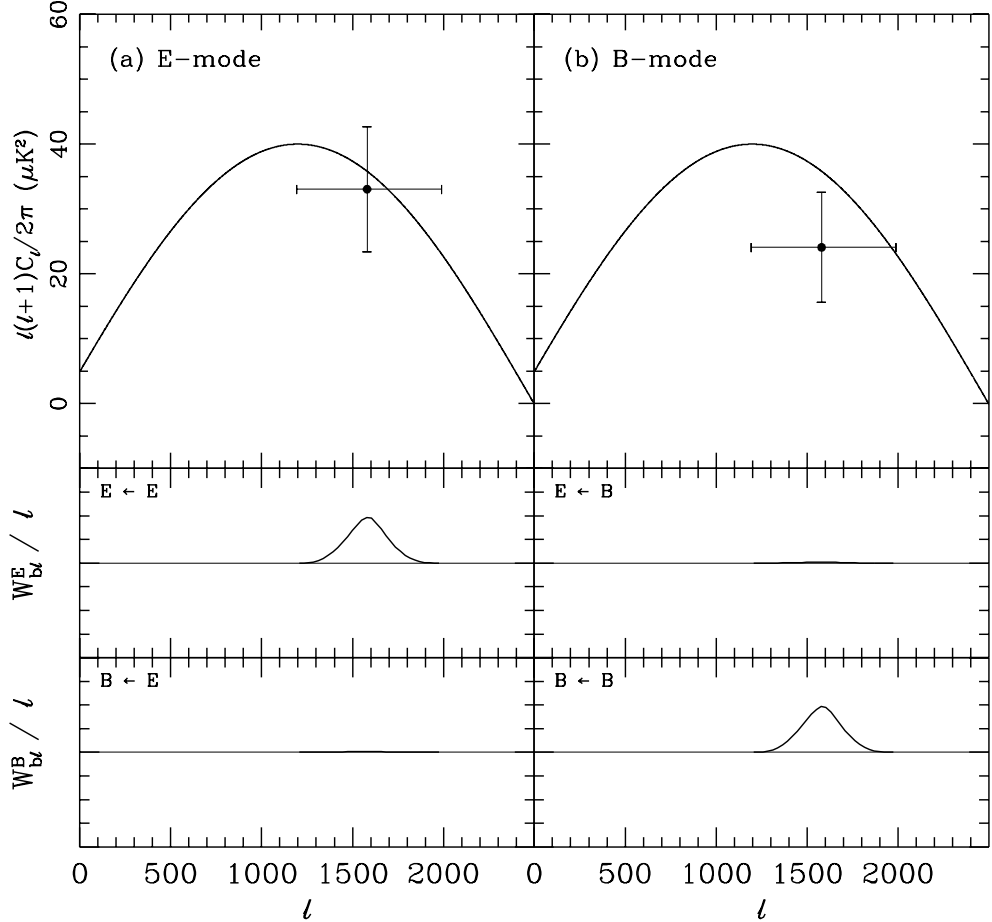}}
\caption{$E$ and $B$ band power estimates (1st row) and the corresponding 
window functions (2nd and 3rd row panels) expected in single-pointing 
observations with the 2-element interferometer ($D=20$ cm and 60 cm separation,
$\nu=95$ GHz). 
Note that the horizontal bar at each band power does not represent the exact 
sensitivity limit in the $\ell$-space, but simply denotes the band width of 
$\Delta\ell \simeq 800$. 
The $\ell$-location of each band power ($\ell_{\rm eff}$) is found from
$\ell_{\rm eff} = \sum_{\ell} \ell (W_{b\ell}/\ell) / \sum_{\ell} (W_{b\ell}/\ell)$,
where the sum is over the values of $\ell$ within the band considered.}
\end{figure}
\clearpage

\begin{figure}
\epsscale{1.5}
\resizebox{\textwidth}{!}{\includegraphics{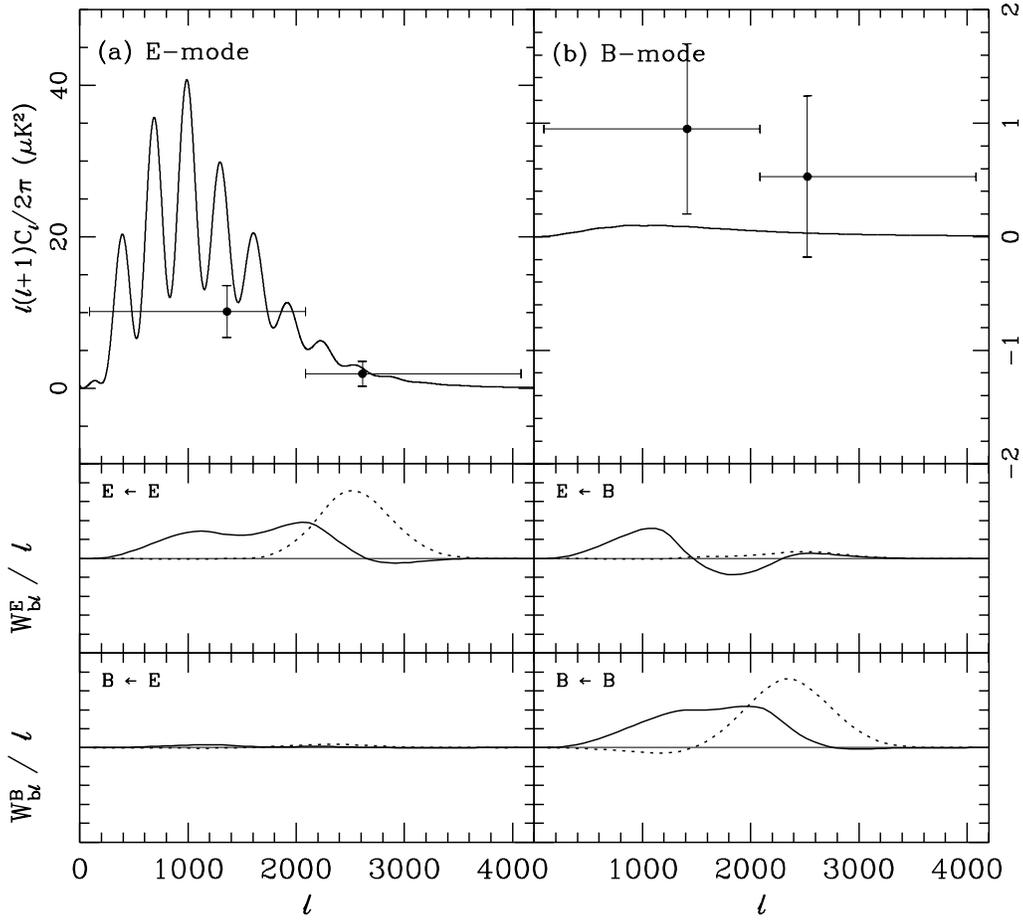}}
\caption{$E$ and lensing induced $B$ band power estimates and the corresponding 
window functions for the $\Lambda$CDM model expected in an one-year 
single-pointing observation with the 7-element AMiBA experiment.}
\end{figure}
\clearpage

\begin{figure}
\epsscale{1.5}
\resizebox{\textwidth}{!}{\includegraphics{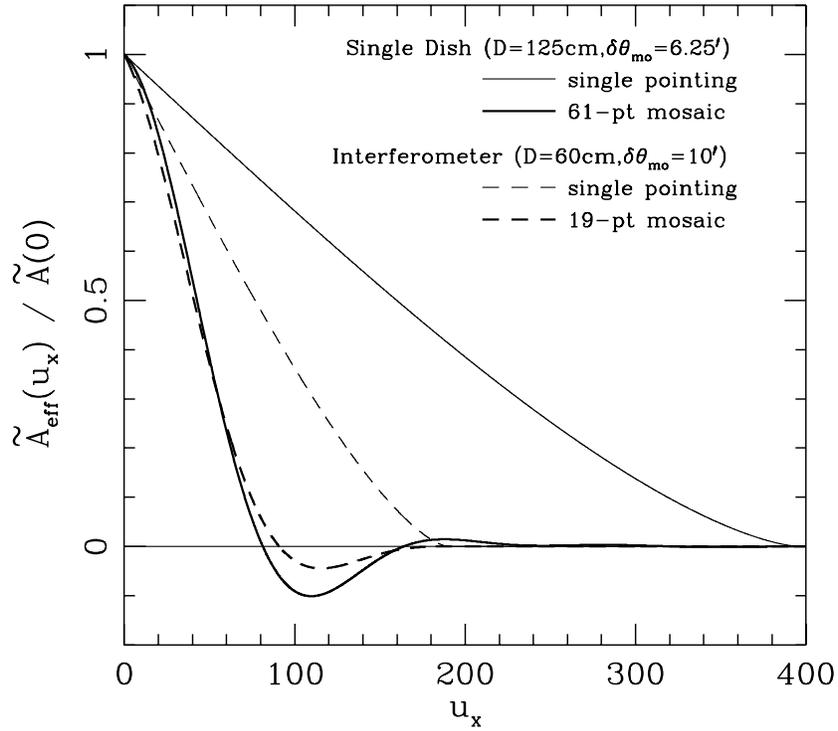}}
\caption{
Effective beam profiles expected in single-pointing (thin) and 
hexagonal mosaicking (thick) observations with a single dish (solid) and a
2-element interferometer with 5 cm separation (dashed), defined as
$\tA_{\rm eff}(\bu)=(1 / N_{\rm mo}) \sum_{m=1}^{N_{\rm mo}} \tA(\bu) 
e^{2\pi i \bu\cdot\by_m}$ (see Park et al. 2003) 
at center frequency $\nu=95$ GHz.
The number of mosaic pointings ($N_{\rm mo}$) and pointing separation 
($\delta\theta_{\rm mo}$) is $(N_{\rm mo},\delta\theta_{\rm mo})=(61,6\farcm25)$
for single-dish, and $(19,10\arcmin)$ for interferometer.}
\end{figure}
\clearpage

\begin{figure}
\epsscale{1.5}
\resizebox{\textwidth}{!}{\includegraphics{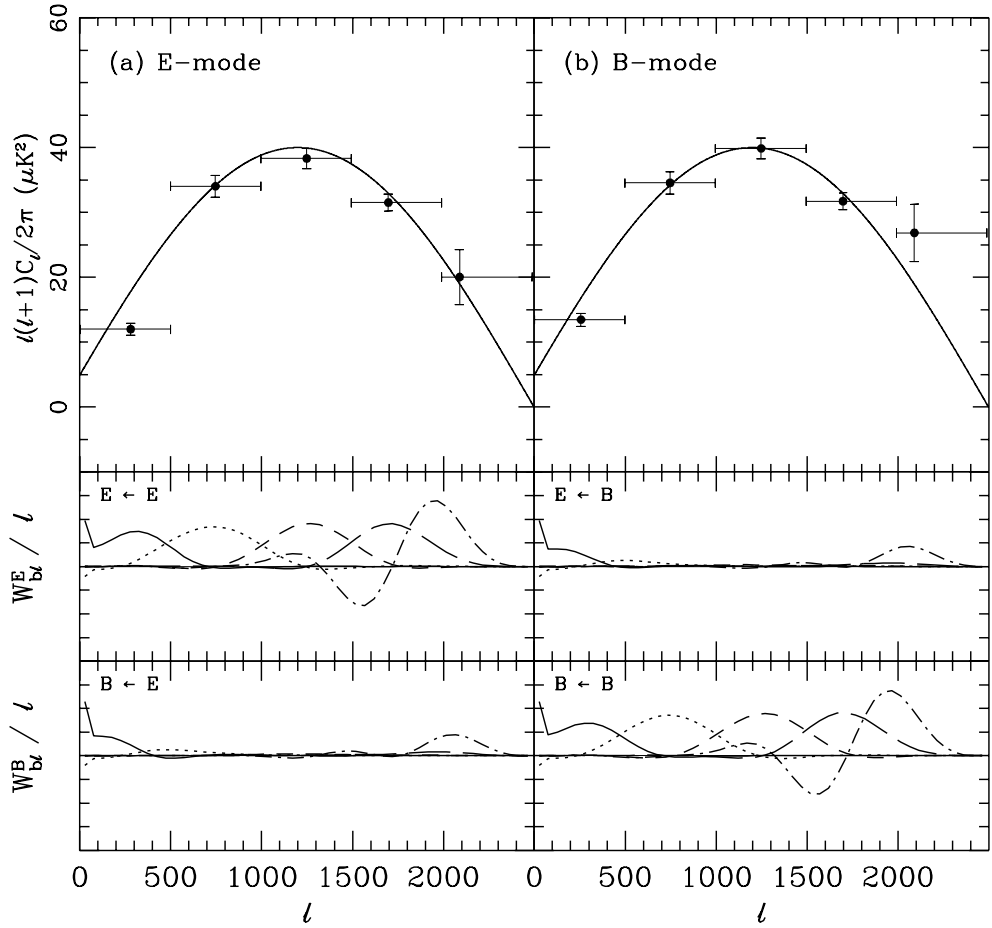}}
\caption{$E$ and $B$ band power estimates (1st row) and the corresponding 
window functions (2nd and 3rd row panels) expected in a 61-pointing hexagonal 
mosaic with the single dish ($D=125$ cm, $\nu=95$ GHz, and 
$\delta\theta_{\rm mo}=6\farcm25$).  
Note that the horizontal bar at each band power simply denotes the band 
width of $\Delta\ell \simeq 500$.}
\end{figure}
\clearpage

\begin{figure}
\epsscale{1.5}
\resizebox{\textwidth}{!}{\includegraphics{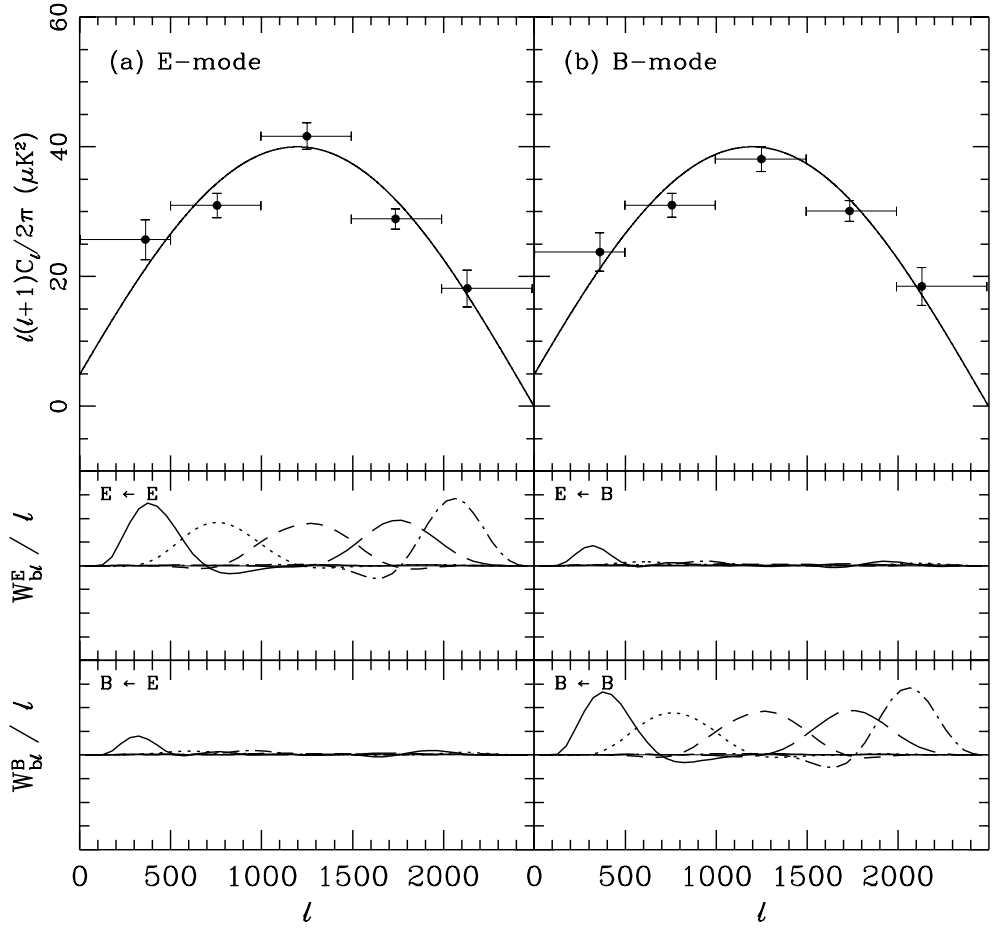}}
\caption{Same as Figure 5 but for interferometric observation, expected 
in a 19-pointing hexagonal mosaic with the 2-element interferometer 
($D=60$ cm and 5 cm separation,
$\nu=95$ GHz, and $\delta\theta_{\rm mo}=10\arcmin$).}
\end{figure}
\clearpage

\begin{figure}
\epsscale{1.5}
\resizebox{\textwidth}{!}{\includegraphics{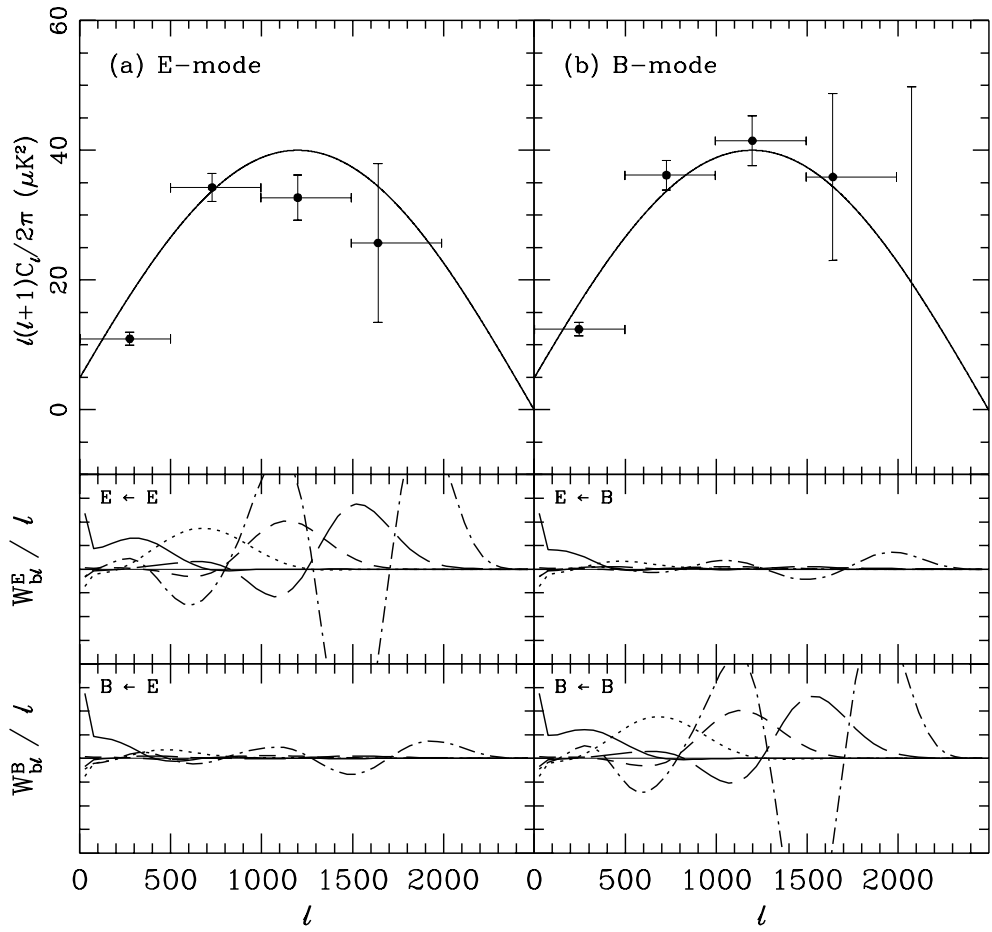}}
\caption{Same as Figure 5 (single-dish experiment) but for realistic 
noise level of 3 $\mu$K per pointing.}
\end{figure}
\clearpage

\begin{figure}
\epsscale{1.5}
\resizebox{\textwidth}{!}{\includegraphics{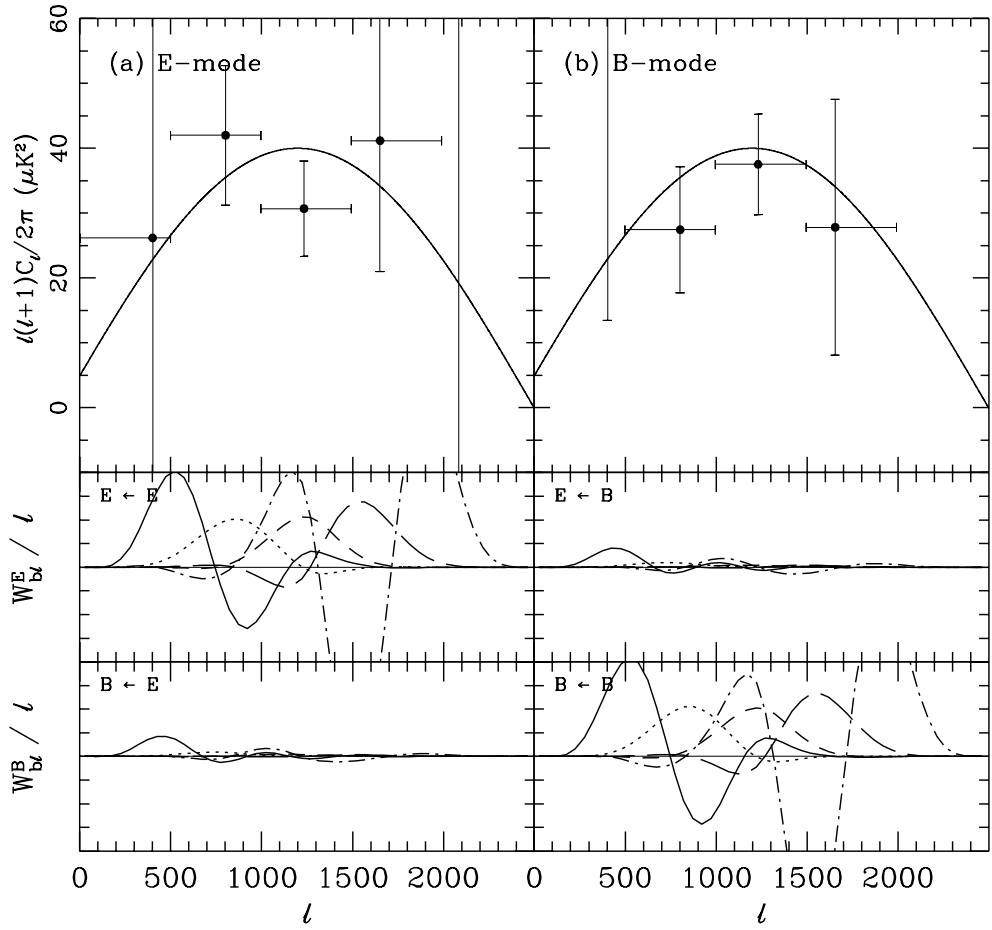}}
\caption{Same as Figure 6 (interferometry) but for realistic noise level 
of 3 $\mu$K per visibility.}
\end{figure}
\clearpage

\begin{figure}
\epsscale{1.5}
\resizebox{\textwidth}{!}{\includegraphics{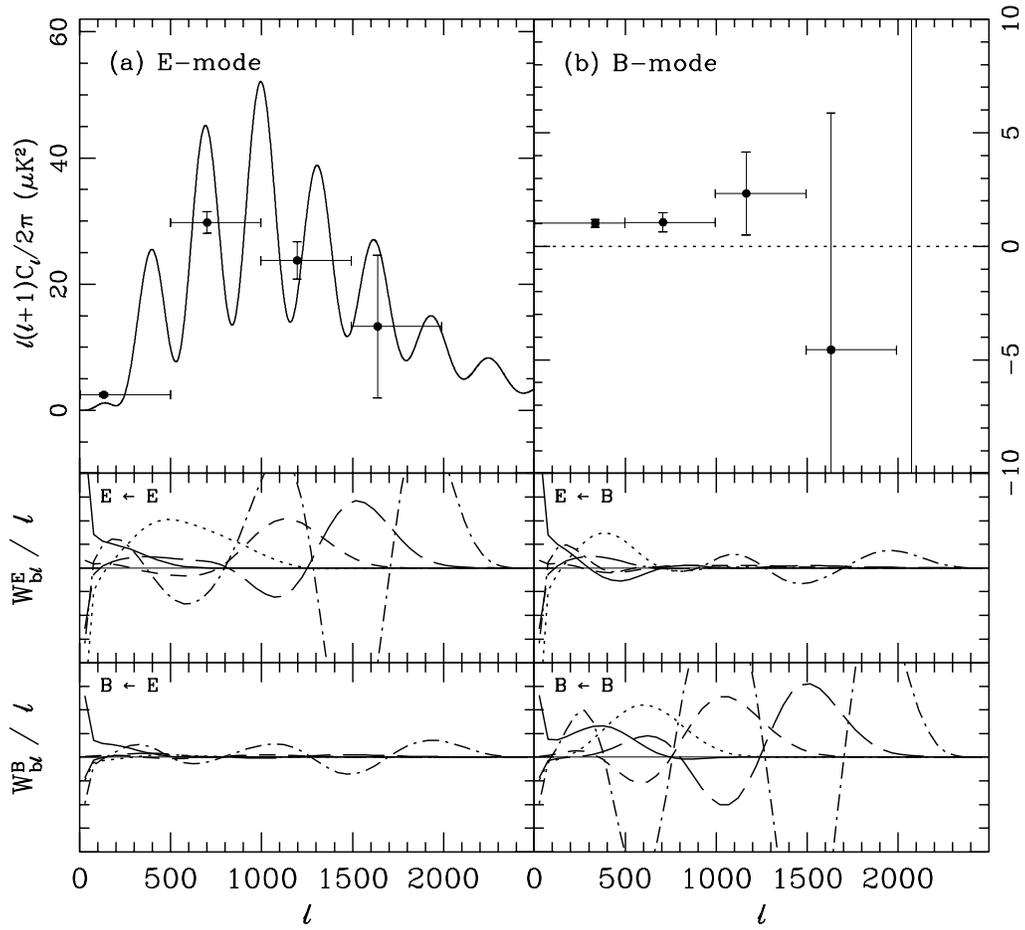}}
\caption{Same as Figure 7 (single-dish experiment) but for realistic 
$\Lambda$CDM power spectrum with zero $B$ mode.}
\end{figure}
\clearpage

\begin{figure}
\epsscale{1.5}
\resizebox{\textwidth}{!}{\includegraphics{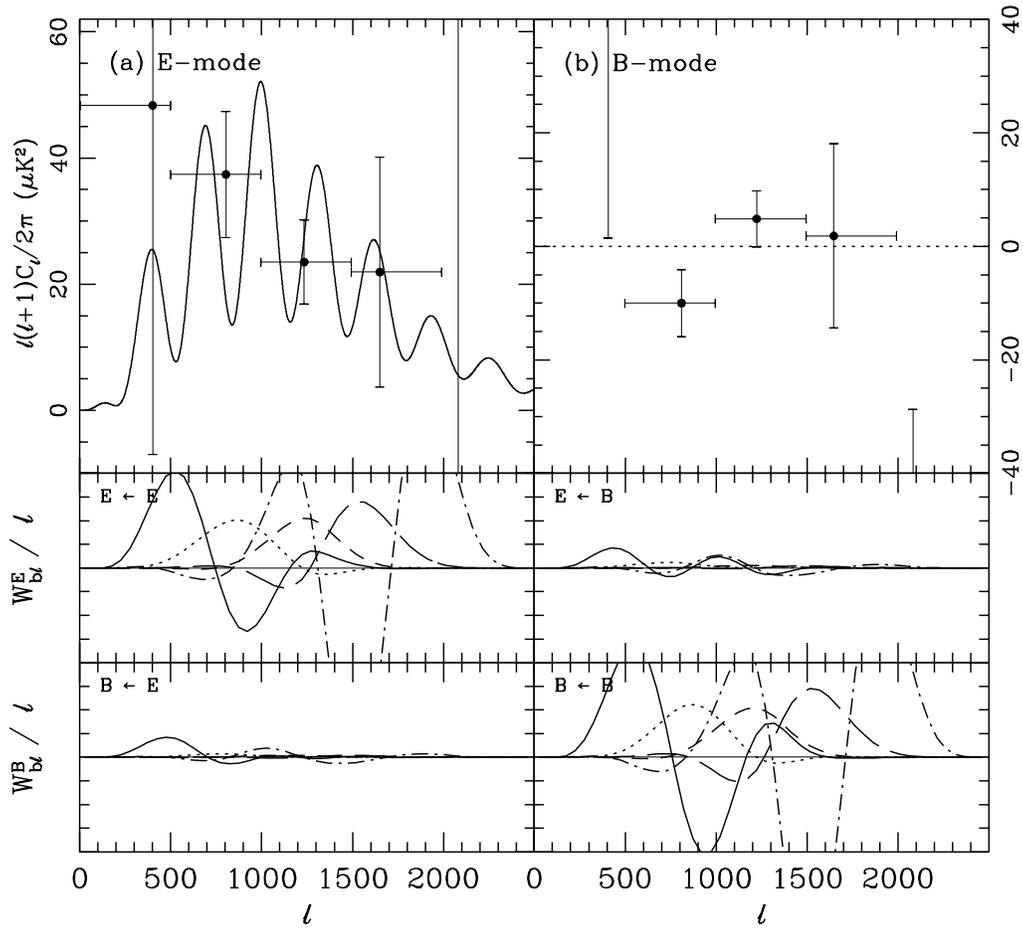}}
\caption{Same as Figure 8 (interferometry) but for realistic $\Lambda$CDM 
power spectrum with zero $B$ mode.}
\end{figure}
\clearpage

\begin{figure}
\epsscale{1.5}
\resizebox{\textwidth}{!}{\includegraphics{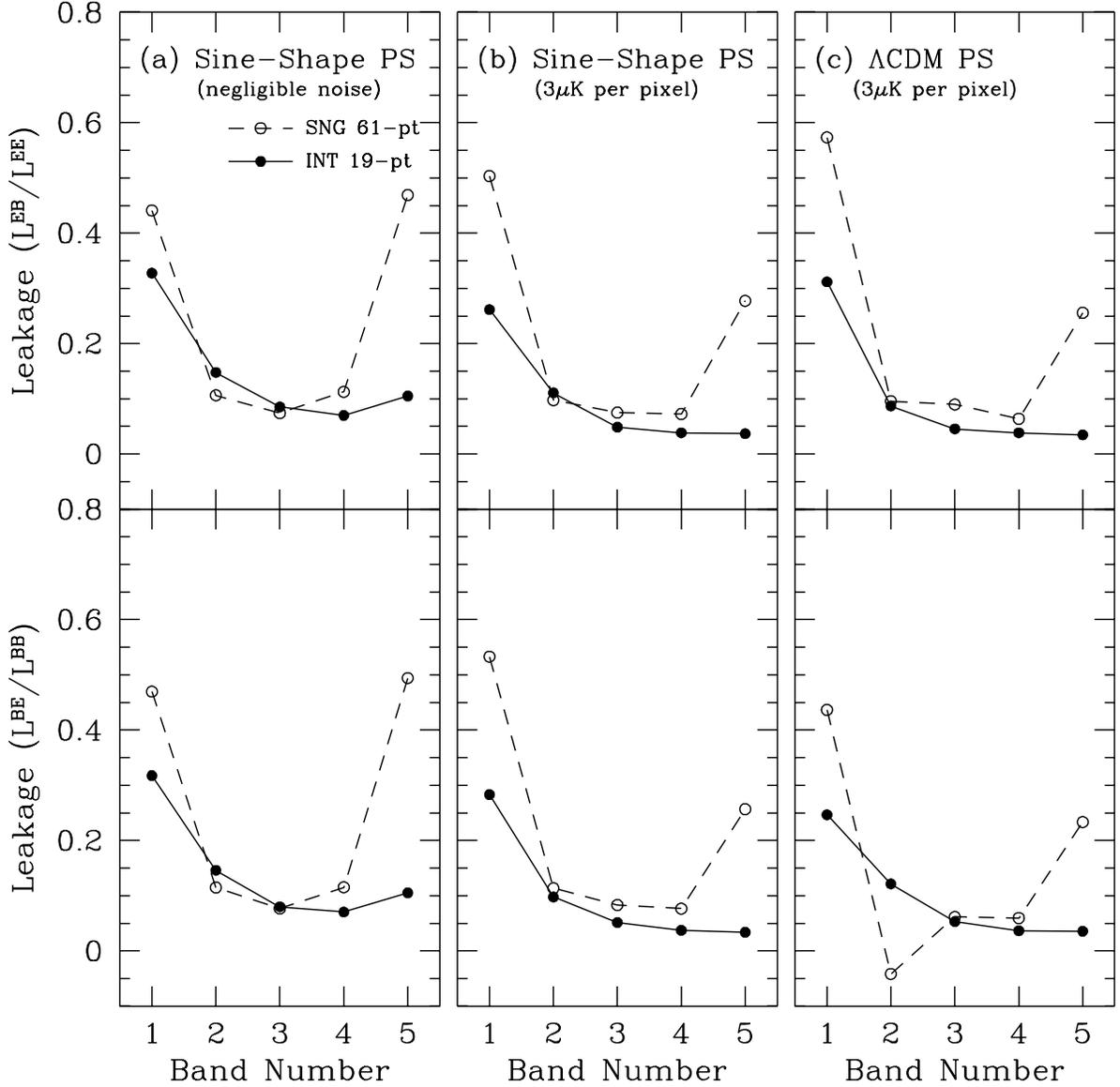}}
\caption{Curves showing leakages of $B$ mode power into $E$-estimates (upper
panels) and of $E$ mode power into $B$-estimates (bottom panels) 
for the sine-shape power spectrum (PS) and the $\Lambda$CDM PS. 
In panel ($a$), the instrument noise is negligible for both the interferometer 
(filled circles with solid curves) and the single dish (open circles with 
dashed curves).
In panels ($b$) and ($c$), realistic noise level (3 $\mu$K per pixel) 
is assumed. The band number denotes the band used in Figures 5--10 with 
increasing $\ell$ order.}
\end{figure}
\clearpage

\end{document}